\documentclass[letterpaper]{jpconf}
\usepackage{graphicx}
\begin{document}
\title{Measurement of ${\cal B}(D^+ \to \mu^+ \nu_{\mu})$ and the pseudoscalar decay constant $f_D$ at CLEO}

\author{Istv\'an Dank\'o (representing the CLEO Collaboration)}

\address{Rensselaer Polytechnic Institute, 110 8th St., Troy, New York 12180, USA}

\ead{idanko@mail.lns.cornell.edu}

\begin{abstract}
The analysis of 60 pb$^{-1}$ data collected on the $\psi$(3770) resonance with the CLEO-c detector has produced the first statistically significant signal for $D^+ \to \mu^+ \nu$ decay and led to the measurement of the decay branching fraction ${\cal B}(D^+ \to \mu^+ \nu) = (3.5 \pm 1.4 \pm 0.6)\times 10^{-4}$ and the pseudoscalar decay constant $f_D = (202 \pm 41 \pm 17)$ MeV.
\end{abstract}

\section{Introduction}

Measurement of the purely leptonic decay of the $D^+$ allows the determination of the pseudoscalar decay constant $f_D$, which can provide important experimental check on Lattice QCD (LQCD) \cite{LQCD} as well as other theoretical predictions \cite{other_theo}, and can help discriminate among these models.
In addition, since LQCD hopes to predict $f_B/f_D$ accurately, measuring $f_D$ would allow a precision determination of the $B$ meson decay constant, $f_B$, which currently can not be measured experimentally.
This would make the precision extraction of the CKM matrix element, $|V_{td}|$, from the $B_d$ meson mixing rate possible.

The branching fraction of the $D^+ \to \ell^+ \nu$ decay in lowest order is given by
\begin{equation}
\label{eq:decay_rate}
{\cal B}(D^+ \to \ell^+ \nu) = \frac{G^2_F}{8\pi}m^2M\left(1-\frac{m^2}{M^2}\right)^2 |V_{cd}|^2 f_D^2 \tau_{D^+},
\end{equation}
where $m$ and $M$ are the mass of the lepton and the $D$ meson, respectively, $G_F$ is the Fermi coupling constant, $V_{cd}$ is the CKM matrix element that quantifies the mixing amplitude for the $c$ and $d$ quark, and $\tau_{D^+}$ is the lifetime of the $D^+$ meson.

\section{Analysis technique}

The data was collected with the CLEO-c detector at the Cornell Electron Storage Ring, a symmetric energy $e^+ e^-$ collider.
In this analysis about 60 pb$^{-1}$ data recorded on the $\psi$(3770) resonance are used.
About half of the time the $\psi$(3770) decays to $D^+ D^-$.
We select these events by reconstructing the $D^-$ meson \footnote{Charge conjugate modes are implied in this analysis.} (tagged $D$) decaying into any one of five modes ($K^+\pi^-\pi^-$, $K^+\pi^-\pi^-\pi^0$, $K_S^0\pi^-$, $K^0_S\pi^-\pi^-\pi^+$, and $K^0_S\pi^-\pi^0$) which represent about 35\% of the total $D$ meson decay rate.
Then we look for $D^+ \to \mu^+ \nu$ signal candidate by searching for an additional single track, presumed to be the muon.

\section{Event selection}

All accepted tracks must be within the fiducial volume of the main drift chamber with a polar angle $|\cos(\theta)|<0.93$ and must come from the $e^+e^-$ interaction point.
Charged kaons and pions are identified by using both the ionization energy loss ($dE/dx$) in the drift chamber and information from the ring imaging Cherenkov (RICH) counters.
Neutral pions are formed from pairs of photon showers detected in the CsI crystal calorimeter and kinematically fitted to the nominal $\pi^0$ mass.
$K^0_S$ candidates are reconstructed from a kinematic fit of a pair of charged pions to a displaced vertex.

\begin{figure}
\begin{center}
\includegraphics*[width=3.4in]{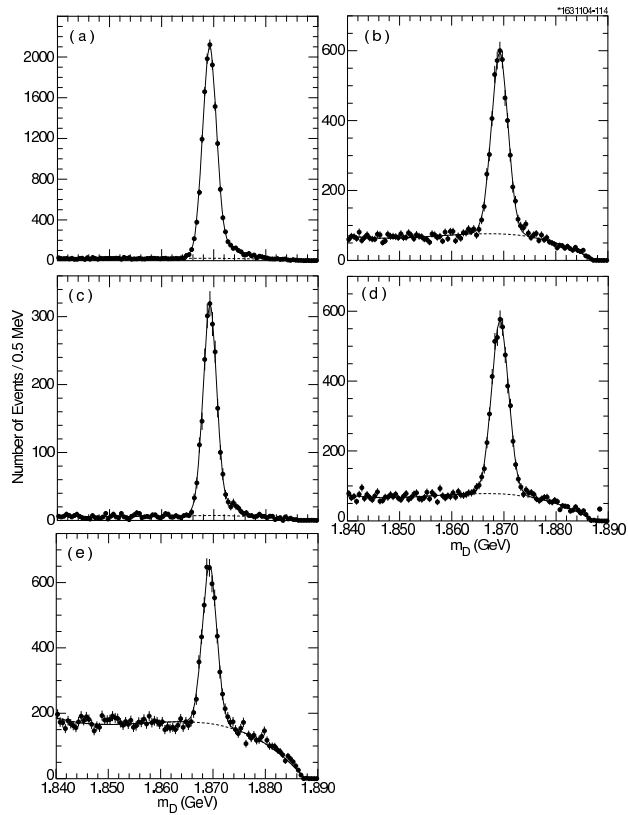}
\end{center}
\caption{Beam constrained mass distributions in data for the five different D tag modes: (a) $D^- \to K^+\pi^-\pi^-$, (b) $D^- \to K^+\pi^-\pi^-\pi^0$, (c) $D^- \to K^0_S\pi^-$, (d) $D^- \to K^0_S\pi^-\pi^-\pi^+$, and (e) $D^- \to K^0_S\pi^-\pi^0$. The solid curves show the sum of the Gaussian signal functions and third order polynomial background functions. The dashed curves indicate the background fits.}
\label{fig:mD_dist}
\end{figure}

The tagged $D$ meson is fully reconstructed by requiring the difference in the energy of the decay products and the beam energy to be less than 20 MeV.
We extract the number of tagged events from fits to the distributions of the beam-constrained mass of the reconstructed $D$ meson defined as $m_D = \sqrt{E^2_{\rm beam} - (\sum \overrightarrow{p})^2}$, where the sum runs over the decay products.
The distributions of $m_D$ for each decay mode are shown in Fig. \ref{fig:mD_dist} where the fit curves are the superposition of Gaussian signal functions and third order polynomial background functions.
The total number of tagged events within $\pm3$ standard deviations of the peak is $28,574 \pm 207 \pm 629$, where the systematic error is estimated from the variation of the signal when using an ARGUS shape function instead of a polynomial function to parametrize the background.

Then the $D^+ \to \mu^+ \nu$ candidates are selected by searching for an additional track with opposite charge to the tagged $D$ meson in the barrel region of the detector ($|\cos(\theta)|<0.81$).
The muon candidate is required to deposit less than 300 MeV energy in the calorimeter, characteristic of a minimum ionizing particle, and not to be consistent with the kaon hypothesis based on the RICH information.
We reject events with more than one additional charged track or if the energy of the largest shower unmatched to a track is more than 250 MeV.

The presence of the neutrino is inferred by requiring that the measured value of the missing mass squared calculated as
\begin{equation}
MM^2 = \left( E_{\rm beam} - E_{\mu^+} \right)^2 - \left( -\overrightarrow{p}_{D^-} - \overrightarrow{p}_{\mu^+} \right)^2,
\end{equation}
be near zero (i.e. the neutrino mass squared), where $\overrightarrow{p}_{D^-}$ is the three-momentum of the fully reconstructed tagged $D^-$ meson.
The resolution of the $MM^2$ is $\sigma = 0.028$ GeV$^2$, which is essentially independent of the tagging decay mode.

\section{Results}

\begin{figure}
\begin{center}
\includegraphics*[width=3.4in]{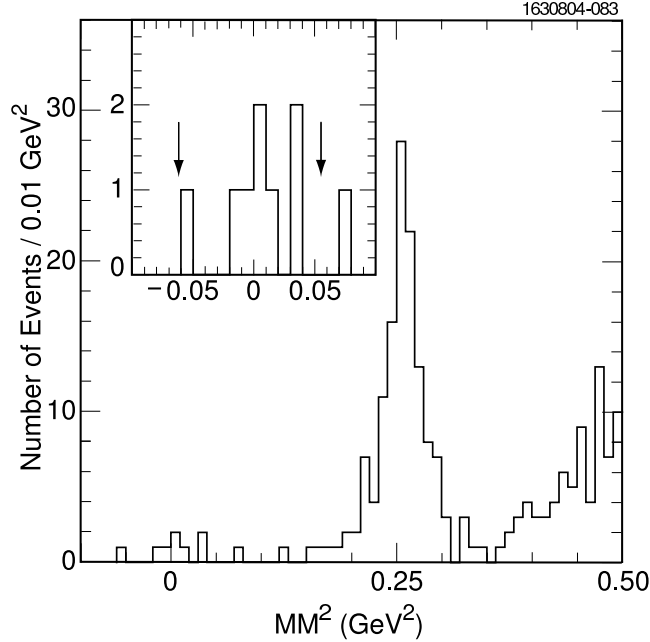}
\end{center}
\caption{$MM^2$ distribution for $D^+ \to \mu^+ \nu$ candidate events. The insert shows the region around zero where the $\pm2\sigma$ signal region is indicated by the arrows.}
\label{fig:MM2}
\end{figure}

Fig. \ref{fig:MM2} shows the $MM^2$ distribution in data for $D^+ \to \mu^+ \nu$ candidate events.
The signal region within $\pm 2 \sigma$ around zero contains 8 events.
The large peak at 0.25 GeV$^2$ is due to the decay $D^+ \to K^0\pi^+$ since many $K^0_L$ escapes detection.

Background can arise from other $D^+$ decay modes, misidentified $D^0 {\bar D}^0$ events and continuum background ($e^+e^- \to q\bar{q}$, $q=u,d,s$).
The likelihood of these events are evaluated using Monte Carlo simulations and the results are summarized in Table \ref{tab:background}. 
The total background is expected to be $1.00 \pm 0.25$ events.
Because of the uncertainties in the Monte Carlo simulation, a 100\% error is assigned to the background estimate.

\begin{table}
  \centering
  \caption{Estimated backgrounds from various sources.}\label{tab:background}
\bigskip
\begin{tabular}[c]{ccc}
  \hline
  \hline
    Background source & ${\cal B}$ (\%) & \# of events \\
  \hline
   $D^+ \to \pi^+ \pi^0$ & $0.13 \pm 0.2$ & $0.31 \pm 0.04$ \\
   $D^+ \to {\bar K}^0 \pi^+$ & $2.77 \pm 0.18$ & $0.06 \pm 0.05$ \\
   $D^+ \to \tau^+ \nu$ & $2.64 \times \cal{B}(D^+ \to \mu^+ \nu)$ & $0.30 \pm 0.07$ \\
   $D^+ \to \pi^0 \mu^+ \nu$ & $0.25 \pm 0.15$ & negligible \\
   $D^0 \bar{D}^0$ & & $0.16 \pm 0.16$ \\
   continuum & & $0.17 \pm 0.17$ \\
  \hline
   total & & $1.0 \pm 0.25$ \\ 
  \hline
  \hline
\end{tabular}
\end{table}

The branching fraction is calculated as ${\cal B}(D^+ \to \mu^+ \nu) = N_{\rm sig}/\varepsilon N_{\rm tag}$, where $N_{sig}=7.0 \pm 2.8$ is the number of background subtracted signal events, $\varepsilon = 0.699$ is the detection efficiency of the decay $D^+ \to \mu^+ \nu$, and $N_{\rm tag}=28,574$ is the number of $D$ tags.
The result is
\begin{equation}
{\cal B}(D^+ \to \mu^+ \nu) = (3.5 \pm 1.4 \pm 0.6) \times 10^{-4}
\end{equation}
The total systematic uncertainty on the branching fraction is $16.4$\%, which is the quadrature sum of the systematic uncertainties in the $D^+ \to \mu^+ \nu$ detection efficiency ($5.3$\%), in the number of $D$ tags ($2.2$\%), and in the background ($15.4$\%).

The decay constant, $f_D$, is obtained from Eq. \ref{eq:decay_rate} by substituting $|V_{cd}| = 0.224$ and $\tau_{D^+} = 1.04$ ps \cite{PDG}
\begin{equation}
f_D = (202 \pm 41 \pm 17) {\rm MeV}.
\end{equation}

\section{Conclusion}

We have reported the recent measurement of the decay branching fraction $D^+ \to \mu^+ \nu$ and the $D$ meson decay constant by the CLEO collaboration \cite{CLEO}.
Our results are considerably smaller, though consistent with previous claim of observation by the BES experiment \cite{BES}.
The experimental value of $f_D$ is consistent with Lattice QCD as well as other model predictions, and the current level of precision is not enough to discriminate among the various models.
The CLEO-c detector continues to take more data on the charm threshold in order to improve the precision of the measurement in the near future.

\ack

We gratefully acknowledge the effort of the CESR staff 
in providing us with excellent luminosity and running conditions.
This work was supported by the National Science Foundation
and the U.S. Department of Energy.

%\medskip

\end{document}